%% file: main.tex
\documentclass[aps,prb,twocolumn,superscriptaddress]{revtex4-2}

\usepackage{amsmath,amssymb}
\usepackage{hyperref}
\usepackage{physics}
\usepackage{graphicx}
\usepackage{color}
\usepackage{float}
\usepackage{bm}
\usepackage{bbm}

\begin{document}

\title{Topological Superconductivity in Altermagnetic Heterostructures on a Honeycomb Lattice}

\author{George McArdle}
\email{george.mcardle@nottingham.ac.uk}
\affiliation{School of Physics and Astronomy, University of Nottingham, Nottingham, NG7 2RD, UK}
\affiliation{Centre for the Mathematics and Theoretical Physics of Quantum Non-Equilibrium Systems, University of Nottingham, Nottingham, NG7 2RD, UK}
\author{Brian Kiraly}
\affiliation{School of Physics and Astronomy, University of Nottingham, Nottingham, NG7 2RD, UK}
\author{Peter Wadley}
\affiliation{School of Physics and Astronomy, University of Nottingham, Nottingham, NG7 2RD, UK}
\author{Adam Gammon-Smith}
\affiliation{School of Physics and Astronomy, University of Nottingham, Nottingham, NG7 2RD, UK}
\affiliation{Centre for the Mathematics and Theoretical Physics of Quantum Non-Equilibrium Systems, University of Nottingham, Nottingham, NG7 2RD, UK}


\date{\today}

\begin{abstract}
Altermagnet-superconductor heterostructures have been shown, in principle, to provide a route towards realising topological superconductivity, and therefore host topologically protected boundary states. In this work we demonstrate that the topological states observed are dependent on the structure of the underlying lattice. By deriving and analysing a model on a honeycomb lattice, we demonstrate that the topological phase diagram has a rich structure containing both chiral edge modes and Majorana corner modes, the latter of which are an indication of higher-order topology. We analyse the effect of disorder on these states and find that whilst the edge modes are robust to a disordered system, any potential observation of the corner modes may be sensitive to the microscopic details. In particular, we show that vacancies can lead to other low energy bound states that may be difficult to distinguish from the corner modes.

\end{abstract}

\maketitle

\section{Introduction}

Topological superconductors (TSCs) are unconventional superconductors that additionally host topologically protected modes either at the boundary of the system or localised at vortices. These protected modes include Majorana modes which have been proposed as a way to realise fault-tolerant quantum computing~\cite{NayakTQCReview, SarmaTQCReview, PachosTQCIntro, TQCWires}.
Despite studies into the unconventional pairing of candidate materials~\cite{TSCMaterials, UPt3, Sr2RuO4}, it has been challenging to verify the presence of topological superconductivity. 

An alternative approach to realise topological superconductors, is to engineer heterostructures typically composed of a superconducting layer with either a semiconductor or a topological insulator.~\cite{LeijnseTSC, SatoTSC, AliceaReview, TI_TSC_Review, StanescuReview, TSCExperimentReview}. This approach has attracted extensive theoretical and experimental attention due to the inherent tunability and the increased availability of the composite parts. The main aim with these setups is to verify the presence of a topological state by confirming the presence of the protected boundary modes and in particular Majorana zero modes (MZMs). It is these zero modes that are of particular interest for proposed implementations of topological quantum computing~\cite{NayakTQCReview, SarmaTQCReview, PachosTQCIntro, TQCWires}, due to their non-Abelian statistics.
Initially introduced in one- and two-dimensional $p$-wave superconductors, MZMs are predicted to exist at the edges of the system or within vortices~\cite{Kitaev_original, Ivanov01}. Despite ongoing efforts however, their conclusive detection remains an open problem. 

Many proposals to realise topological superconductivity in engineered systems involve semiconductor-superconductor heterostructures. In these, an $s$-wave superconductor is placed in proximity to a semiconductor with strong spin-orbit coupling. In the presence of a magnetic field it creates an effective $p$-wave spinless superconductor~\cite{SemiconductorPlatform, AliceaSemiconductors, OregMBS, LutchynSemiconductors, MZMsSolidState, MZMPhaseTransition, RobustSMC}. However, there are still two major obstacles. The first is the presence of disorder in the system, which can result in low-energy states that are difficult to distinguish from MZMs~\cite{AndreevBoundStates, ZeroBiasPeaks, DasSarmaExptReview} and the second is that the magnetic field required to break the time-reversal symmetry can also destroy the superconducting state (see~\cite{AliceaReview} for a discussion of this). Recent research has suggested that the use of an altermagnet will assist in addressing these issues~\cite{Hughes_Altermagnet}.

Altermagnets are a recently discovered but potentially abundant form of magnetism which have attracted interest in a variety of areas, including spintronics~\cite{AMSpintronics1, AMSpintronics2}, anomalous Hall transport \cite{AHE1, AHE2, AHE3} and various aspects of superconductivity~\cite{MajoranaTwist, Hughes_Altermagnet, CornerModes, CornerModes2, AMSC1, AMSC2, AMSCBeenakker, AMTopSC}. They are characterised by sublattices of opposite spin that are related not by translation or inversion but by rotation~\cite{EmergingAMReview, AMReview2, AMReview3, AMReview}. This results in spin-split bands without requiring relativistic effects such as spin-orbit coupling. The spin-split bands have a momentum dependence that resembles the standard $d$-, $g$-, or $i$-wave symmetries.  Additionally, altermagnets also break time-reversal symmetry but with net zero magnetisation. It is this latter property that is crucial for proposed realisations of topological superconductivity since the magnetic field, which would suppress the superconducting gap, is no longer required.

In particular, two-dimensional altermagnet-superconductor heterostructures can, in principle, host both Majorana edge modes~\cite{Hughes_Altermagnet, MajoranaTwist} and corner modes \cite{CornerModes, CornerModes2}. These reflect different topological phases, with the former being indicative of a first-order topological superconductor, whereas the corner modes are a sign of higher-order topology~\cite{HOTIs, BBH_Science, SecondOrderTSCs, Khalaf2018}. In addition, recent experimental developments in controlling and observing altermagnetic systems~\cite{PeterBrianPaper} further motivate the need to assess the feasibility of altermagnet heterostructures as a way of realising topological superconductivity.

In this work, we explore the altermagnet-superconductor heterostructures on a honeycomb lattice. This extends the existing proposals for a square lattice. We identify that such a change in the lattice structure can have an impact on the topological states seen. To demonstrate this we show how the Chern number varies with different parameters, highlighting the distinct topological regimes consisting of edge modes and corner modes. Furthermore, we explore the effect of disorder on these modes, which is a key consideration in any future experimental implementation.

\section{Results}

\begin{figure}
    \centering
    \includegraphics[width=1\linewidth]{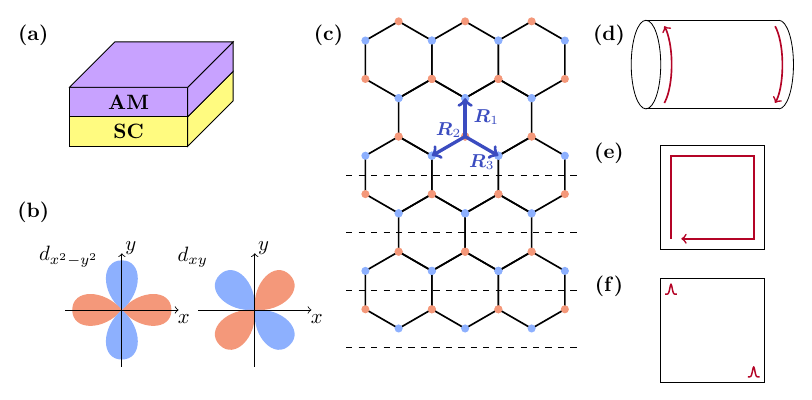}
    \caption{The altermagnet-superconducting heterostructure is shown in (a). The altermagnet is $d$-wave so that the exchange terms follow the symmetries depicted in (b). Furthermore, this work considers a honeycomb lattice, see (c), with sublattices $A$ (red) and $B$ (blue). The size of the lattice is $N_x \times N_y$, with $N_x=7$ being the number of sites in a row (rows separated by dashed lines) and $N_y=6$ is the number of rows. The vectors, ${\bm R_j}$, define the nearest neighbours. This system hosts various topological regimes, including chiral edge modes which can be seen using periodic boundary conditions (i.e. on a cylinder as in (d)) or in real space, as in (e). Zero-energy modes localised to the corners of the sample can also be observed as depicted in (f).}
    \label{fig:HexagonalLattice}
\end{figure}

We model the altermagnet-superconducting heterostructure (Fig.~\ref{fig:HexagonalLattice}(a)) using an effective two-dimensional tight-binding model. In this model, we consider a two-dimensional $d$-wave altermagnet with Rashba spin-orbit coupling. The $d$-wave nature means that the magnitude of the exchange coupling term, which is proportional to ${\bm \sigma}\cdot {\bm J}$, is modulated by the two possible $d$-wave symmetries, $d_{x^2-y^2}$ and $d_{xy}$ as depicted in Fig.~\ref{fig:HexagonalLattice}(b). Here, ${\bm \sigma}$ is a vector of Pauli matrices acting on the spin degree of freedom and ${\bm J}$ is the exchange coupling vector which points in the direction of the Néel vector. Further to this, our model is on a honeycomb lattice, which is bipartite with two distinct sites in the unit cell, $A$ and $B$ (see Fig.~\ref{fig:HexagonalLattice}(c)) with an energy difference of $2M$.

In order to include superconductivity, we couple an $s$-wave superconductor to the altermagnet. This induces an $s$-wave pairing term in our model via the proximity effect and allows us to use the effective two-dimensional model outlined here. We provide further details of the model, including the Hamiltonian, in the Methods section. Furthermore, we consider different boundary conditions (BCs) for our model. Periodic BCs in both the vertical and horizontal directions allows us to study the bulk physics, whereas with periodic BCs in one direction we explore edge modes on a cylinder (Fig.~\ref{fig:HexagonalLattice}(d)). Finally we look at open boundary conditions in both directions, allowing us to identify edge modes and corner modes, as sketched in Fig.~\ref{fig:HexagonalLattice}(e) and (f) respectively.

\subsection{Bulk Behaviour and Edge Modes}\label{sec:edge}

\begin{figure}
    \centering
    \includegraphics[width=1\linewidth]{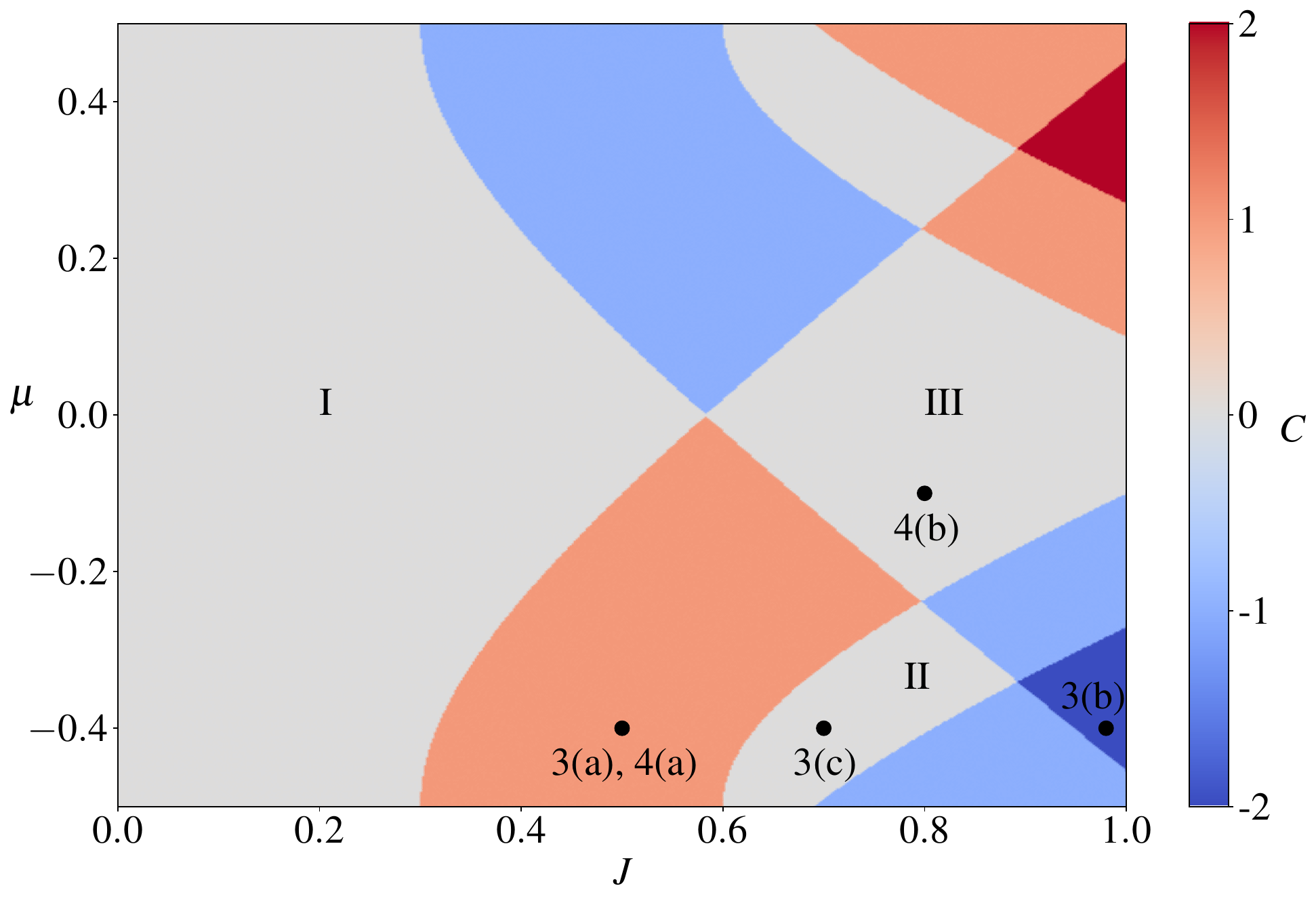}
    \caption{The Chern number as a function of the magnitude of the coupling vector $J$ and chemical potential $\mu$ for the $d_{x^2-y^2}$ symmetry ($\alpha_1=1$, $\alpha_2=0$ in the Hamiltonian). The phase diagram, plotted for $500 \times 500$ uniformly spaced data points, contains a trivial region with no edge modes (region I) as well as regions with edge modes and corner modes. The regions containing topological boundary states can be divided into two classes; first regions with  $C\neq 0$ host $|C|$ chiral edge modes representing a first-order TSC, and second, regions with $C=0$. This can be due to either counter-propagating edge modes (region II) or a higher-order TSC (HOTSC) which hosts corner modes (region III). The additional labels correspond to other figures that show the topological states. The parameters used are $M=0$, $\lambda=\Delta=0.3$. The Néel vector is assumed to point in the $z$-direction.}
    \label{fig:HexagonalChern}
\end{figure}

In order to analyse the topological behaviour of this model, we first need to identify the regimes where non-trivial topological effects occur. To achieve this we calculate the Chern number for different parameters to obtain a phase diagram. The Chern number is invariant in each topological phase and only changes during topological phase transitions when a band gap in the bulk spectrum closes~\cite{BernevigHughes}. We compute the Chern number, $C$, using the Fukui-Hatsugai-Suzuki method~\cite{FHS_Method} (see Methods section for further details). From this we are able to analyse the bulk behaviour of the system as well as any chiral edge modes due to the bulk-boundary correspondence (see, for example~\cite{SatoTSC, TI_review_KaneHasan}). In Fig.~\ref{fig:HexagonalChern}, we present a phase diagram showing how the Chern number varies with the exchange coupling, $J$, and the chemical potential, $\mu$. Although there are several other parameters in our model, tuning these two illustrates the essential physics of the system. Furthermore, the qualitative behaviour of the results is similar across a range of values for both the spin-orbit coupling constant, $\lambda$, and the induced superconducting pairing, $\Delta$. We will discuss the effects of varying the other parameters throughout the remainder of this work.

The set of parameters we have used to obtain Fig.~\ref{fig:HexagonalChern} are motivated using insights from the square lattice results \cite{Hughes_Altermagnet, CornerModes, CornerModes2}. For example, the exchange coupling vector ${\bm J}$, which points in the direction of the Néel vector, is taken to be completely of out-of-plane. This is because, although there can be some in-plane component whilst maintaining the same qualitative behaviour, as the out-of-plane component decreases the bulk gap closes and the Chern number becomes ill-defined. 

Additionally, we introduce asymmetry in the nearest neighbour hopping amplitudes and unless specified otherwise we have $t_1=t_3=1$ and $t_2=0.8$ ($t_i$ is the hopping amplitude in the direction ${\bm R_i}$). In the case of the square lattice, a similar asymmetry ensured the system behaved as a strong topological superconductor rather than a weak one. For the honeycomb lattice, the asymmetry between the sublattices is not required to obtain a strong TSC with chiral edge modes, but is instrumental in the realisation of a higher-order topological phase hosting corner modes. The asymmetry between the $A$ and $B$ sublattices is achieved by this asymmetric hopping, and so for simplicity we set the energy offset between the two sublattices as $M=0$. A finite $M$, with symmetric hopping can also break the symmetry between sublattices and may be relevant for experimental systems. However, making this distinction does not significantly impact the regions with chiral edge modes in Fig.~\ref{fig:HexagonalChern}.

The phase diagram shown in Fig.~\ref{fig:HexagonalChern} exhibits a rich structure for $d_{x^2-y^2}$ symmetry (see Fig.~\ref{fig:HexagonalLattice}(b)) of the exchange coupling. Similar to the square lattice \cite{Hughes_Altermagnet}, there is a large region of the parameter space (region I) where the both the bulk and edge spectra are fully gapped and the system is in a topologically trivial phase. In this region $C=0$ and there are no chiral edge modes or corner modes. 

\begin{figure}[H]
    \centering
    \includegraphics[width=1\linewidth]{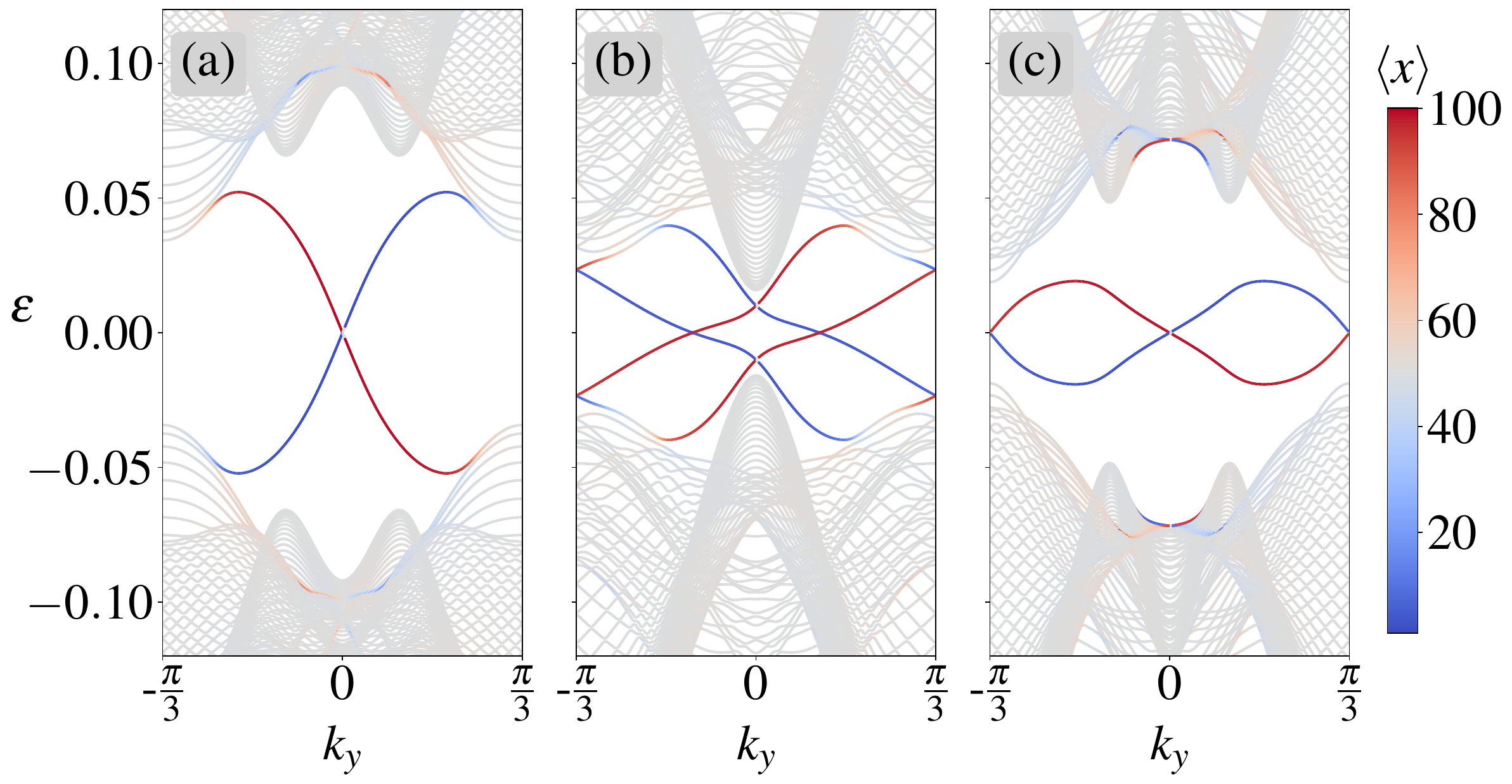}
    \caption{The energy spectrum of modes on a cylinder geometry. The cylinder is periodic in the $y$-direction and has $100$ sites in the $x$ direction. The average position of the modes along the cylinder, $\langle x \rangle$ is denoted by the colour. (a) corresponds to a regime where $C=1$ where there is a single chiral edge mode per edge. In (b), where $C=-2$ there are 2 edge modes per edge with opposite chirality to (a), and (c) corresponds to region II in Fig.~\ref{fig:HexagonalChern} where $C=0$. In all plots, $\mu=-0.4$. The values of $J$ used are: (a) $J=0.5$, and (b) $J=0.98$, and (c) $J = 0.7$.}
    \label{fig:HexagonalEdges}
\end{figure}

However, there are also multiple regimes where the Chern number is non-trivial and the system hosts topological edge modes. For example, there are regions with $|C|=1$, which corresponds to one chiral edge mode per edge. We demonstrate this by assuming periodic boundary conditions in the $y$-direction only, to create a cylinder geometry. The resulting energy spectrum is shown in Fig.~\ref{fig:HexagonalEdges}(a), where each edge hosts a chiral edge mode. This is consistent with a single chiral edge mode in a system with open boundary conditions, as is shown in Fig.~\ref{fig:CornerModes}(a). We note that an opposite sign of the Chern number simply reflects the opposite chirality in the edge mode.

The presence of a non-zero Chern number and therefore chiral edge modes is similar to what was predicted for the square lattice~\cite{Hughes_Altermagnet} and is an indication of a strong topological superconductor.  Further to this, there is the appearance of a regime with $|C|=2$. Such a regime is not predicted for the square lattice and corresponds to two co-propagating chiral modes per edge, as shown in Fig.~\ref{fig:HexagonalEdges}(b). Similar behaviour has in some circumstances been linked to the presence of a quantum anomalous Hall (QAH) state when the pairing, $\Delta$, is tuned to zero~\cite{Hughes_QAH_TSC, Magnetic_QAH_TSC}. Although the anomalous Hall effect has been studied in altermagnets~\cite{AHE_AM1,AHE_AM2,AHE_AM3,AHE_AM4} with some proposals extending to the QAH effect~\cite{QAH_altermagnet}, we suspect this is not the cause here. While the origin of the $|C|=2$ regime is not fully understood, we believe we can rule out the relation to QAH effect due to the absence of topologically protected chiral edge modes in the limit $\Delta\rightarrow0$. The presence of these edge modes is a necessary signature of any QAH regime.

In addition to the regimes addressed so far, there are two other phases in Fig.~\ref{fig:HexagonalChern}. Both region II and region III appear trivial through calculation of the Chern number alone. Region II, however, has counter-propagating edge modes (Fig.~\ref{fig:HexagonalEdges}(c)). Similar modes have been found to not backscatter in topological superconductor setups due to the presence of an additional symmetry, such as a mirror symmetry \cite{MirrorChern}. Further research into the symmetries of this model could lead to insights into the origins of these modes in this setup.

\subsection{Corner Modes}\label{sec:corner}

\begin{figure}
    \centering
    \includegraphics[width=\linewidth]{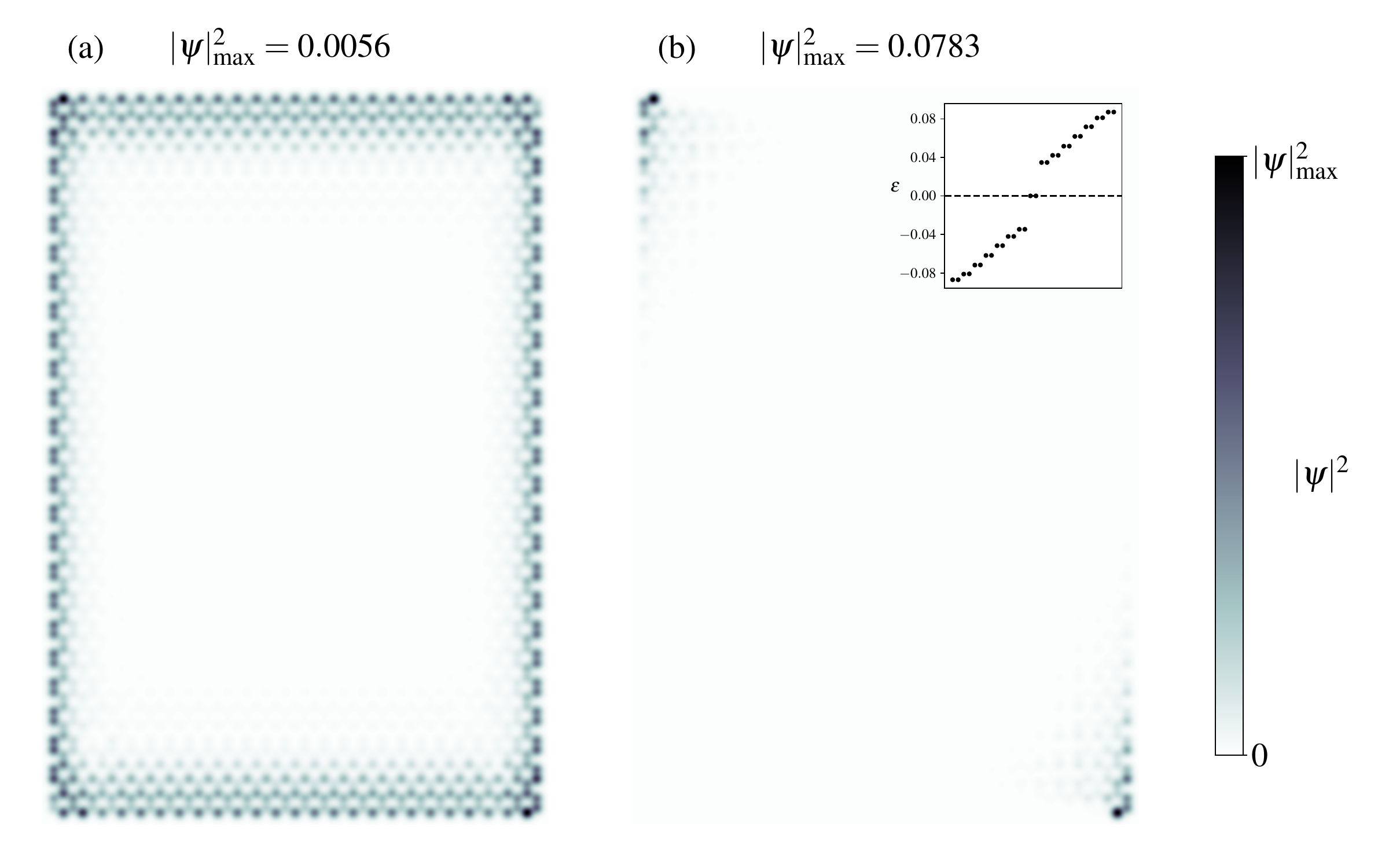}
    \caption{The eigenstates with the lowest absolute energy plotted as a sum of Gaussian functions centred on each site weighted by the real-space on-site probability $|\psi(x)|^2 = |\langle x | \psi\rangle|^2$. The size of the lattice is $51 \times 50$ sites. The parameters shown are:  (a) $J=0.5$  $\mu=-0.4$, resulting in Chern number $|C|=1$ and therefore a gapless chiral edge mode; (b) $J=0.8$, $\mu=-0.1$, which corresponds to region III. Here $|C|=0$ and there are zero-energy (see inset) modes localised to the corners of the sample.}
    \label{fig:CornerModes}
\end{figure}

In region III of Fig.~\ref{fig:HexagonalChern} zero-energy Majorana corner modes are present as we show in Fig.~\ref{fig:CornerModes}(b), indicating a second-order topological superconducting phase. A similar result is seen for the square lattice when an additional component such as a topological insulator with gapless helical edge modes~\cite{CornerModes, CornerModes2} or a $p$-wave superconductor~\cite{AltermagnetBraiding} is added. Here, however the sublattice degree of freedom acts in a comparable way to the orbital degree of freedom usually provided by the topological insulator and so the simple heterostructure is sufficient.

The higher-order topology and the Majorana corner modes arise when the edges can themselves be described by one-dimensional topological Hamiltonians~\cite{BBH_Science, BBH_PRB, HOTIs, HighT_MCMs, HighT_MCMs2}. In particular, when two adjacent edges have a gapped Hamiltonian but with a mass term of opposite sign a zero energy mode must reside at the corner. The corner modes are then robust providing the edge or the bulk gap doesn't close. In this instance, the transition away from the corner mode regime involves a bulk gap closing and so there are challenges in constructing an analytical edge theory. However within region III we find that for $\mu \neq 0$ the bulk and edge gaps remain and the corner modes are robust, but for $\mu=0$ the edge gap closes and no corner modes are present.

\begin{figure}
    \centering
    \includegraphics[width=1\linewidth]{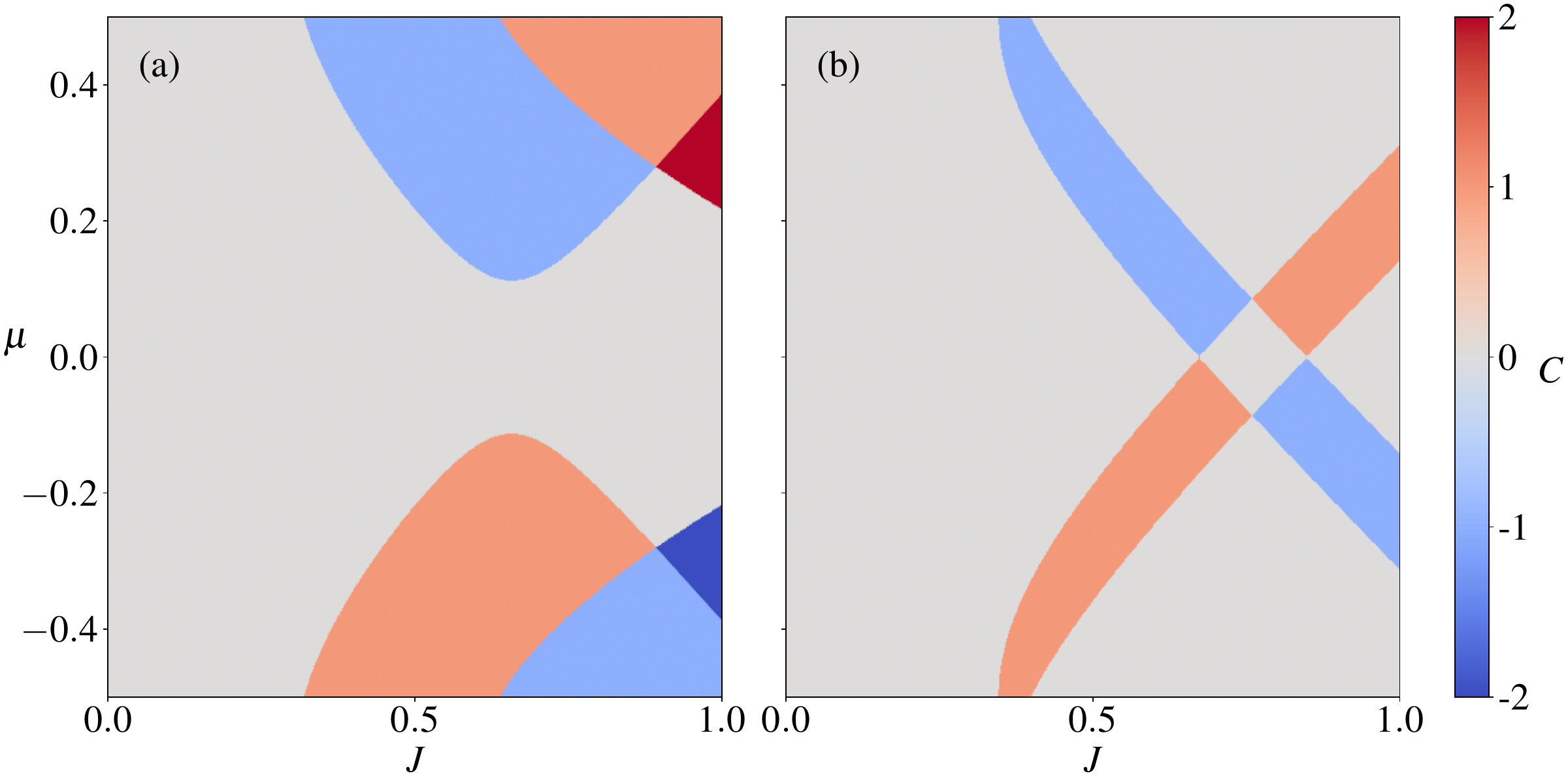}
    \caption{(a): The phase diagram for symmetric hopping, that is $t_1=t_2=t_3=1$, with the sublattice asymmetry given by $M=0.1$. The orbital symmetry used is $d_{x^2-y^2}$. The lattice can still host zero-energy corner modes, for example at $\mu=-0.15, J=0.85$. (b): The phase diagram for the $d_{xy}$-orbital ($\alpha_1=0, \alpha_2=1$ in the Hamiltonian) is shown for $M=0$ and asymmetric hopping. The diamond at the centre of the plot contains corner modes for $\mu \neq 0$. The top, right, and bottom $C=0$ regions contain gapped-out edge modes.}
    \label{fig:M_d_phase}.
\end{figure}

In addition to requiring a non-zero chemical potential, which can be controlled experimentally using a gate voltage, the realisation of this higher-order TSC also requires an asymmetry between the $A$ and $B$ sublattices. Here, this is ensured by the asymmetry in the hopping amplitudes. However, an alternative way to achieve this is by having symmetric hopping but an energy difference between the sublattices. In our model this can be accomplished by $M\neq 0$. As we show in Fig.~\ref{fig:M_d_phase}(a) for $M=0.1$, the phase diagram is unaffected for large regions of parameter space. The most notable changes are the disappearance of region II where there were counter-propagating modes and also there is now no bulk gap closure between the regions I and III. However, the system still hosts localized corner modes in certain regions of the phase diagram where $C=0$, for example, when $\mu=-0.15, J=0.85$. Unlike when the asymmetry of the sublattices was caused by the different hopping strengths, here the transition to the regime hosting corner modes involves the closing of a gap in the edge spectrum, but no closing of the bulk gap. In order to identify the parameters at which the corner modes occur a more detailed analysis is required, through either an exploration of where the gaps in the edge spectra occur, or a calculation of higher-order topological invariants \cite{BBH_Science, BBH_PRB, HOTIs}. 
 
Further to the consideration of finite $M$, another factor that may be relevant to experimental proposals is the effect of the $d_{xy}$-symmetry in the exchange coupling. This was found not to result in topological physics in the square lattice, however, its affect on the honeycomb lattice is more subtle. Indeed if we consider the situation where the sublattice asymmetry is caused by the different onsite energies, i.e. $M\neq0$, then the $d_{xy}$-orbital alone does not result in interesting topological physics, with the Chern number being trivial across the entire phase diagram considered. Despite this, its inclusion does alter the phase diagram when both orbitals are included, in particular is the reappearance of a region with counter-propagating edge modes (see the Supplementary Material). 

In contrast to this, when the asymmetry is contained in the hopping amplitudes, the $d_{xy}$-orbital itself results in non-trivial topology, as shown in Fig.~\ref{fig:M_d_phase}(b). Although the phase diagram does not contain as many non-trivial regions as for $d_{x^2-y^2}$, there are still signatures of both first-order and higher-order topology. There are regions containing chiral edge modes ($C\neq 0$) and we note that the `diamond' shape in the centre of the figure also hosts corner modes for $\mu \neq 0$, exactly as occurs for the $d_{x^2-y^2}$-orbital. When both orbitals are present, the resulting phase diagram is a combination of the two individual orbitals (see the Supplementary Material) and therefore it is clear that their interplay can affect the topology seen. In the development of more accurate models, particularly those that are tailored to specific materials, it is imperative that the effect of these orbitals is appropriately accounted for.

\subsection{Disorder}

In any experimental system, the presence of disorder is unavoidable and so it is important to consider its impact. Here we look at the effect vacancies have on the edge and corner modes. We consider different `structures' of the random disorder configurations, such as a disordered boundary, random vacancies throughout the system (with varying amounts of disorder), and clusters of vacancies which can be on the boundary or in the bulk - the latter of which corresponds to a hole in the sample. 

\begin{figure}
    \centering
    \includegraphics[width=\linewidth]{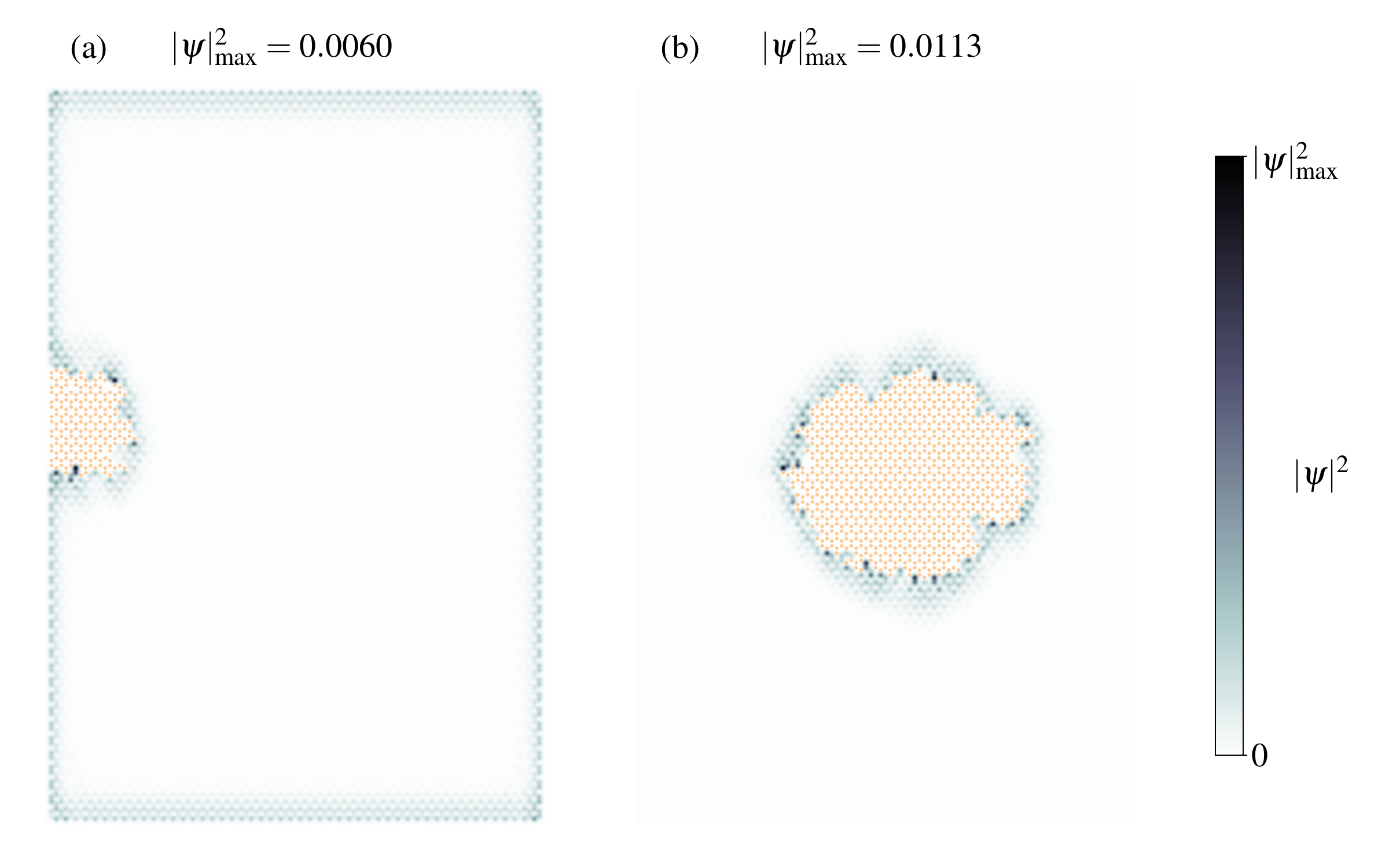}
    \caption{The lowest absolute energy eigenstate in the edge mode regime (second lowest in (b)) as a sum of Gaussians each weighted by the on-site probability, in the presence of disorder. The system size with no vacancies is $101\times 100$ sites. The edge modes are robust to a disordered boundary, including if there are a cluster of vacancies (small orange diamonds) on the boundary as in (a). The outer edge modes are also robust to bulk disorder, including a hole in the sample. If the hole is sufficiently large such that it behaves as another edge, then an additional inner edge mode appears at this boundary, as shown in (b). In both plots $J=0.5$ and $\mu=-0.4$.}
    \label{fig:EdgeModeDisorder}
\end{figure}

For a disordered boundary, the edge modes persist due to their topological nature. This is most clearly seen if we consider a cluster of vacancies on the boundary, as in Fig.~\ref{fig:EdgeModeDisorder}(a). Here the edge mode clearly survives and travels around the missing sites, as expected. This outer edge mode similarly survives a random cluster of vacancies in the centre of the sample. Furthermore, if this hole is sufficiently large (larger than the correlation length of the edge modes), then it acts as a boundary to the system and an additional inner edge mode circulates the edge of the hole. We show this additional edge mode in Fig.~\ref{fig:EdgeModeDisorder}(b). Similarly the corner modes are robust to clusters of vacancies (see Supplementary Material).

\begin{figure}
    \centering
    \includegraphics[width=1\linewidth]{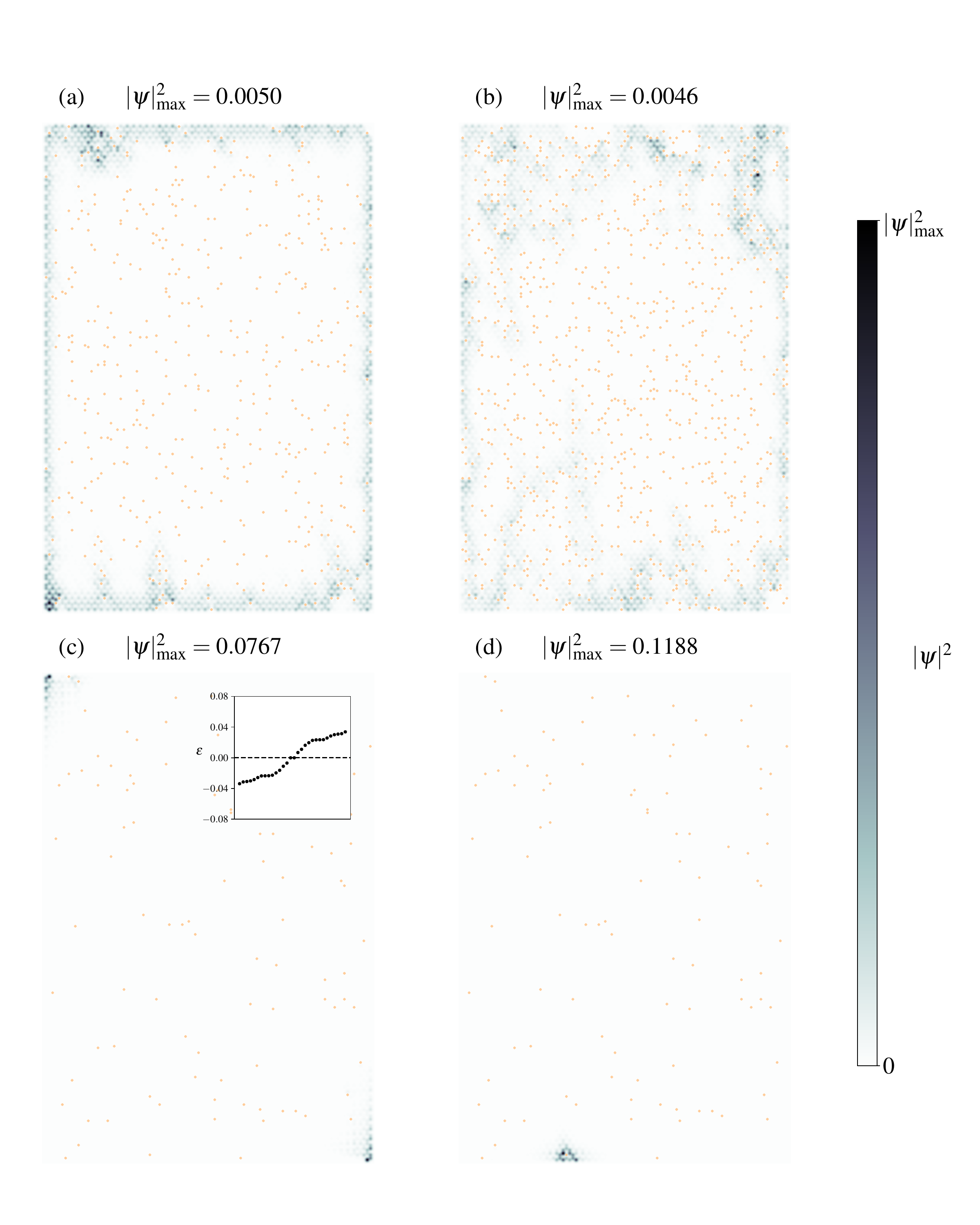}
    \caption{A sum of Gaussians weighted by the on-site probability for edge ((a),(b): $J=0.5, \mu=-0.4$) and corner ((c), (d): $J=0.8, \mu=-0.1$) modes in the presence of a random disorder configuration (shown by small orange diamonds) for a $101\times 100$ site system. (a), (b), (c) are the eigenstates with the smallest absolute energy and (d) has the second smallest absolute energy. The edge modes are robust (although less well defined) to $5\%$ disorder as in (a) and become less localised as the disorder is increased further with $10\%$ of sites being vacancies in (b). In (c), it is clear that the corner modes can survive at $1\%$ disorder but there may additional modes close to zero energy (see inset for the spectrum). These modes are not localised to the corners, but are pinned to impurities as shown in (d).}
    \label{fig:random_disorder}.
\end{figure}

We now consider random disorder where a percentage of random sites in the lattice are missing. When $1\%$ of the sites are missing, the edge modes survive and rearrange themselves to avoid any vacancies located at the boundary. The edge modes are still present for a disorder level of $5\%$, although they are less well defined (see Fig.~\ref{fig:random_disorder}(a)) and become spread across the system as disorder is further increased, as we show for a disorder level of $10\%$ in Fig.~\ref{fig:random_disorder}(b). This demonstrates that whilst the edge modes are robust to disorder, a relatively clean system is still necessary in order to be able to identify them.

On the other hand, the presence of disorder in the system has a significant impact on the potential observation of corner modes. In this case, even at low levels of disorder, there could be difficulties in their detection. In Fig.~\ref{fig:random_disorder}(c) we show that the corner modes survive at $1\%$ disorder. However, as can be seen in the inset, there are also now additional states close to zero energy. These are not found in the corners of the system (see Fig.~\ref{fig:random_disorder}(d)) and could be experimentally difficult to distinguish from the corner modes. Although here we have focused on a single disorder realisation, the appearance of low energy modes pinned to the impurities occurs for other realisations and disorder structures, and can also affect the spatial extent of the corner modes (see the Supplementary Material). It is clear from this that the microscopic details or the disorder will have an impact on any potential observation of the corner modes and observation of robust edge modes may be more conclusive.

\section{Discussion}

To summarise, we have shown that a clean altermagnet-superconductor heterostructure can exhibit both first-order and higher-order topology if the underlying lattice has a honeycomb structure. The first-order topology is characterised by a non-zero Chern number and the presence of chiral edge modes, while the higher-order topology has gapped edge spectra and Majorana corner modes. We have explored the robustness of the boundary states to different types of disordered structures, ranging from clusters of vacancies to random vacancies throughout the sample. We find that the edge modes are robust to all of these structures, and that the  corner modes are robust to a variety of disorder realisations. However, the corner modes are more sensitive to the particular microscopic details. In particular the appearance of other low-energy bound states may make the corner modes difficult to observe experimentally.

Our work also indicates that while the underlying lattice affects the type of topological states that can be observed, there are a broad range of parameters over which they can be seen. This flexibility is significant for experimental efforts to detect Majorana modes, as various parameters can be tuned (for example the chemical potential using a gate voltage) to access different topological regimes. Furthermore, both the edge modes and corner modes can potentially be used for the realisation of topological quantum computation~\cite{TQCWires, EdgeModeBraiding, EdgeModesTQC}, with corner modes previously being a particular focus in altermagnetic systems~\cite{AltermagnetBraiding}. A more detailed investigation into the control over topological boundary states would allow for a greater understanding of the potential of these heterostructures in quantum technologies.

\section{Methods}\label{sec:methods}

\subsection{Model}

We model the altermagnet using a two-dimensional, free-fermion model on a bipartite lattice. By denoting the annihilation (creation) operators for the sites of type $A$ (located at ${\bm r_i}$)  and $B$ (at ${\bm r_i}+{\bm R_j}$) as $a_i, b_{i+j} \hspace{5pt} (a_i^\dagger, b_{i+j}^\dagger)$ respectively, the 2d tight-binding Hamiltonian for the altermagnet is

\begin{widetext}
\begin{multline}\label{H_AM}
    H_{\rm AM} = \sum_{\langle i,j\rangle} \sum_{\sigma = \uparrow, \downarrow}\frac{t_j}{2} \left(b^\dagger_{i+j, \sigma} a_{i, \sigma} + a^\dagger_{i, \sigma} b_{i+j, \sigma} \right) + (M-\mu)\sum_{i \in A}\sum_{\sigma = \uparrow, \downarrow} a^\dagger_{i, \sigma} a_{i, \sigma} - (M+\mu)\sum_{i \in B} \sum_{\sigma = \uparrow, \downarrow} b^\dagger_{i, \sigma} b_{i, \sigma} \\ + \frac{1}{2} \sum_{\langle i, j \rangle} \sum_{\alpha} f_\alpha(\varphi_j) \left(b^\dagger_{i+j} ({\bm \sigma} \cdot {\bm J})a_i + a^\dagger_i ({\bm \sigma} \cdot {\bm J})b_{i+j} \right).
\end{multline}
\end{widetext}
The hopping strengths in each direction are $t_j$, the chemical potential is $\mu$ and the two sublattices have onsite energies that differ by $2M$. The final term describes the exchange coupling with ${\bm \sigma}$ being a three-dimensional vector of Pauli matrices acting in the spin space and ${\bm J}$ is the exchange coupling vector that points in the direction of the Néel vector. In this work, we focus on $d$-wave altermagnets and the relevant symmetry is encapsulated in the function 

\begin{equation}\label{f_def}
    f_\alpha(\varphi_j) = \begin{cases}
        \alpha_1 \cos(2\varphi_j), \hspace{10pt} \textrm{for the } d_{x^2-y^2} \textrm{ orbital},  \\ \alpha_2 \sin(2\varphi_j), \hspace{10pt} \textrm{for the } d_{xy} \textrm{ orbital}, 
    \end{cases}
\end{equation}
where $\alpha_{1,2}$ are coupling strengths of each orbital and $\varphi_j$ is the angle between the $x$-axis and ${\bm R_j}$. Although we have focused on a $d$-wave exchange term here, the model can be generalised to other symmetries, such as $g$-wave.

We also include spin-orbit coupling in the altermagnet via a standard nearest neighbour Rashba term, with coupling constant $\lambda$,

\begin{equation}\label{H_SOC}
    H_{\rm SOC} =  -\frac{i\lambda}{2} \sum_{\langle i,j \rangle} \left(b^\dagger_{i+j}({\bm \sigma} \times {\bm R_j})_z a_i - a^\dagger_i({\bm \sigma} \times {\bm R_j})_z b_{i+j}\right).
\end{equation}
Here, the operators contain both spin components, $a_i = (a_{i, \uparrow}, a_{i, \downarrow})^{\rm T}$ and the Pauli matrices, $\sigma_i$, act in this spin subspace.

To model the heterostructure, we couple the altermagnet to a conventional $s$-wave superconductor via the proximity effect. Using the same procedure as for semiconductor-superconductor heterostructures - see, for example \cite{AliceaReview}), we find that an $s$-wave pairing term is induced in our effective 2d model for the heterostructure.
This gives the contribution to the Hamiltonian,

\begin{multline}\label{H_Delta}
        H_\Delta = -\Delta \sum_{i \in A} \left(a^\dagger_{i, \uparrow} a^\dagger_{i, \downarrow} + a_{i, \downarrow} a_{i, \uparrow} \right) \\-\Delta \sum_{i \in B} \left(b^\dagger_{i, \uparrow} b^\dagger_{i, \downarrow} + b_{i, \downarrow} b_{i, \uparrow} \right),
\end{multline}
where $\Delta$ is the induced pairing strength (which differs from the pairing strength in the bare superconductor \cite{AliceaReview}). The full Hamiltonian is given by $H = H_{\rm AM} + H_{\rm SOC} + H_\Delta$.

For the honeycomb lattice, depicted in Fig.~\ref{fig:HexagonalLattice}, the nearest vector neighbours are ${\bm R_1} = \hat{\bm y}$,  ${\bm R_2} = -\tfrac{\sqrt 3}{2}\hat{\bm x}-\tfrac{1}{2}\hat{\bm y}$, and ${\bm R_3} = \tfrac{\sqrt 3}{2}\hat{\bm x}-\tfrac{1}{2}\hat{\bm y}$, where we have assumed a unit lattice spacing. To correctly account for the impact of each of the $d$-orbitals in Eq.~\ref{f_def}, the angles of the nearest neighbour vectors measured from the $x$-axis are needed. The angle for the $j$-th nearest neighbour is $\varphi_j = \tfrac{\pi}{2}+(j-1)\tfrac{2\pi}{3}$.
Inserting these nearest neighbours into the Hamiltonian, we obtain the real space Hamiltonian for a honeycomb lattice with two inequivalent sublattices. The bulk Hamiltonian can be obtained by assuming periodic boundary conditions in both directions and taking the Fourier transform. The full expression for the momentum space Hamiltonian in Bogoliubov-de-Gennes (BdG) form is provided in the Supplementary Material. Here we wish to draw attention to the symmetries of the bulk Hamiltonian. Once in BdG form, terms in the Hamiltonian can be expressed as a Kronecker product of Pauli matrices acting on the three subspaces of the system; particle-hole ($\tau_i$), spin ($\sigma_i)$ and sublattice ($\gamma_i$). It is easy to verify that the Hamiltonian obeys particle-hole symmetry by confirming that $\tau_x H^*(-{\bm k})\tau_x= -H({\bm k})$. A similar condition on the spin subspace is necessary to check for time-reversal symmetry $H({\bm k})=\sigma_yH^*(-{\bm k})\sigma_y$. However in this model, TRS is broken by the altermagnet. Specifically, in BdG form the altermagnetic exchange terms become
\begin{multline}
    J_x^c({\bm k}) \tau_z \otimes \sigma_x \otimes \gamma_x - J_x^s({\bm k}) \tau_z \otimes \sigma_x \otimes \gamma_y \\+ J_y^c({\bm k}) \tau_0 \otimes \sigma_y \otimes \gamma_x - J_y^s({\bm k}) \tau_0 \otimes \sigma_y \otimes \gamma_y \\+ J_z^c({\bm k}) \tau_z \otimes \sigma_z \otimes \gamma_x - J_z^s({\bm k}) \tau_z \otimes \sigma_z \otimes \gamma_y.
\end{multline}
The $J$'s here have the properties that $J_i^c(-{\bm k}) = J_x^c({\bm k})$ and $J_i^s(-{\bm k}) = - J_x^s({\bm k})$ with the full expressions given in the Supplementary Material. It can be easily verified that these break TRS. Therefore the system is a topological superconductor in Altland-Zirnbauer class D, and so the Chern number is the topological invariant of interest \cite{AltlandZirnbauer, TopologyPeriodicTable}.

\subsection{Chern Number}

The Chern number is calculated by integrating the Berry curvature over the Brillouin zone via the standard Fukui-Hatsugai-Suzuki method \cite{FHS_Method}. Here we briefly outline the general procedure used in this work. First, using a $501 \times 501$ grid we discretise the Brillouin zone, defined by the reciprocal lattice vectors ${\bm b_1} = \left(\frac{2\pi}{\sqrt{3}}, \frac{2\pi}{3} \right), \hspace{5pt} {\bm b_2}= \left(-\frac{2\pi}{\sqrt{3}}, \frac{2\pi}{3} \right)$. Then by computing the product of overlaps between eigenvectors around a plaquette, we can calculate the flux through a plaquette. In this step it is important to account for the multiband nature of the problem as there are four occupied bands here (since the Hamiltonian has eight bands and is particle-hole symmetric) and so the overlaps of the eigenstates form a matrix whose determinant should be taken instead of just the overlap. The final step is to take the sum over all plaquettes. 

\subsection{Disorder}

We model disorder as a set of impurities (located at sites $\alpha$) with the Hamiltonian
\begin{equation}
    H_{\rm imp}=\sum_\alpha V c_\alpha^\dagger c_\alpha.
\end{equation}
A vacancy corresponds to the limit $V\rightarrow\infty$. In order to implement this computationally, we decouple the vacancy from all the neighbouring sites (so there is no hopping to or from the site) and apply a large onsite potential $(1\times 10^6)$.

\begin{acknowledgements}
A.G-S. and G.M. were supported by UK Research and Innovation (UKRI) under the UK government's Horizon Europe funding guarantee [grant number EP/Y036069/1]. P.W. acknowledges support from the ERC Advanced Grant no. 101095925. and the EPSRC grant EP/V031201/1. B.K. was supported by the EPSRC grant EP/Y023250/1.
\end{acknowledgements}

\appendix
\section*{Supplementary Material}

\input{additional_content.tex}

\bibliography{bibliography.bib}

\end{document}

%% file: additional_content.tex
\section*{The Hamiltonian on the Honeycomb Lattice}\label{appendix:Hex_Ham}
Here, we will give further details to the bipartite Hamiltonian proposed in Eqs.~(1)-(4) and give explicit expressions for our model on the honeycomb lattice. We highlight the symmetries of the Hamiltonian and therefore demonstrate that the Chern number is the relevant invariant quantity. 

In order to calculate the Chern number using the standard Fukui-Hatsugai-Suzuki method \cite{FHS_Method} we require the Hamiltonian in momentum space. To get this we assume periodic boundary conditions in both directions and perform the Fourier transform. To do so in a consistent manner it is important for us to note that the normalisation of the Dirac-Delta function is defined such that,
\begin{equation}
    \sum_{i \in A} {\rm e}^{i({\bm k}-{\bm k'})\cdot {\bm r_i}} = \sum_{i \in B} {\rm e}^{i({\bm k}-{\bm k'})\cdot {\bm r_i}} = \frac{1}{2}\delta({\bm k}-{\bm k'}).
\end{equation}

This allows us to easily perform the Fourier transform, which will eventually enable us to explore the symmetries of the Hamiltonian. 

Performing the full Fourier transform, leads to the full Hamiltonian for the 2d heterostructure becoming
\begin{widetext}
    \begin{multline}\label{H_k_space}
        H = \frac{1}{2}\sum_{\bm k}\Bigg(\sum_{\sigma, j} \frac{t_j}{2}\left(\cos({\bm k}\cdot{\bm R_j})\left[b^\dagger_{{\bm k}, \sigma}a_{{\bm k}, \sigma} + a^\dagger_{{\bm k}, \sigma}b_{{\bm k}, \sigma}\right]-i\sin({\bm k}\cdot{\bm R_j})\left[b^\dagger_{{\bm k}, \sigma}a_{{\bm k}, \sigma} - a^\dagger_{{\bm k}, \sigma}b_{{\bm k}, \sigma}\right]\right)  \\ + \frac{1}{2}\sum_{j, \alpha}f_\alpha(\varphi_j)\left(\cos({\bm k}\cdot{\bm R_j})\left[b^\dagger_{{\bm k}}({\bm \sigma}\cdot{\bm J})a_{{\bm k}} + a^\dagger_{{\bm k}}({\bm \sigma}\cdot{\bm J})b_{{\bm k}}\right] -i\sin({\bm k}\cdot{\bm R_j})\left[b^\dagger_{{\bm k}}({\bm \sigma}\cdot{\bm J})a_{{\bm k}} - a^\dagger_{{\bm k}}({\bm \sigma}\cdot{\bm J})b_{{\bm k}}\right]\right) \\ -\frac{i\lambda}{2}\sum_j \left(\cos({\bm k}\cdot{\bm R_j})\left[b^\dagger_{{\bm k}}({\bm \sigma} \times {\bm J})_za_{{\bm k}} - a^\dagger_{{\bm k}}({\bm \sigma}\times{\bm J})_zb_{{\bm k}}\right] - i \sin({\bm k}\cdot{\bm R_j})\left[b^\dagger_{{\bm k}}({\bm \sigma} \times {\bm J})_za_{{\bm k}} + a^\dagger_{{\bm k}}({\bm \sigma}\times{\bm J})_zb_{{\bm k}}\right]\right) \\+ \sum_\sigma \left((M-\mu)a^\dagger_{{\bm k}, \sigma}a_{{\bm k},\sigma} - (M+\mu)b^\dagger_{{\bm k}, \sigma}b_{{\bm k},\sigma} \right)- \Delta\left(a^\dagger_{{\bm k}, \uparrow}a^\dagger_{-{\bm k}, \downarrow} + a_{-{\bm k}, \downarrow}a_{{\bm k}, \uparrow}+b^\dagger_{{\bm k}, \uparrow}b^\dagger_{-{\bm k}, \downarrow} + b_{-{\bm k}, \downarrow}b_{{\bm k}, \uparrow} \right)\Bigg).
    \end{multline}
\end{widetext}

We additionally note, that a similar procedure is used to obtain the model in the cylinder geometry, where periodic boundary conditions are assumed to be in one direction only.

\subsection{Honeycomb Lattice}

Using the result in Eq.~(\ref{H_k_space}), we can easily verify the square lattice result.  Due to the equivalence of the sublattices in the square lattice case, we can set $M=0$ and define a new fermionic operator $c_{{\bm k},\sigma}\equiv a_{{\bm k},\sigma}= b_{{\bm k},\sigma}$ to recover the result obtained in \cite{Hughes_Altermagnet}.

Unlike the square lattice, there aren't any clear simplifications for the honeycomb lattice so we simply substitute the nearest neighbours and their corresponding angles into Eq.~(\ref{H_k_space}). The vectors to nearest neighbours are shown in Fig.~1 and are given by ${\bm R_1} = \hat{\bm y}$ $(\varphi_1 = \pi/2)$,  ${\bm R_2} = -\tfrac{\sqrt 3}{2}\hat{\bm x}-\tfrac{1}{2}\hat{\bm y}$ $(\varphi_2 = 7\pi/6)$, and ${\bm R_3} = \tfrac{\sqrt 3}{2}\hat{\bm x}-\tfrac{1}{2}\hat{\bm y}$ $(\varphi_3 = 11\pi/6)$, where we have made clear in the corresponding angle in brackets. This leads to the Hamiltonian in Bogoliubov-de-Gennes form (with $\Psi_{\bm k}=(a_{{\bm k}, \uparrow} \hspace{5pt} b_{{\bm k}, \uparrow} \hspace{5pt} a_{{\bm k}, \downarrow} \hspace{5pt} b_{{\bm k}, \downarrow} \hspace{5pt} a^\dagger_{-{\bm k}, \uparrow} \hspace{5pt} b^\dagger_{-{\bm k}, \uparrow} \hspace{5pt} a^\dagger_{-{\bm k}, \downarrow} \hspace{5pt} b^\dagger_{-{\bm k}, \downarrow})^{\rm T}$ due to the extra sublattice degree of freedom)

\begin{widetext}
    \begin{multline}
        H_{\bm k} = M \tau_z \otimes \sigma_0 \otimes \gamma_z - \mu \tau_z \otimes \sigma_0 \otimes \gamma_0 + t_c({\bm k}) \tau_z \otimes \sigma_0 \otimes \gamma_x - t_s({\bm k}) \tau_z \otimes \sigma_0 \otimes \gamma_y + J_x^c({\bm k}) \tau_z \otimes \sigma_x \otimes \gamma_x - J_x^s({\bm k}) \tau_z \otimes \sigma_x \otimes \gamma_y \\ + J_y^c({\bm k}) \tau_0 \otimes \sigma_y \otimes \gamma_x - J_y^s({\bm k}) \tau_0 \otimes \sigma_y \otimes \gamma_y + J_z^c({\bm k}) \tau_z \otimes \sigma_z \otimes \gamma_x - J_z^s({\bm k}) \tau_z \otimes \sigma_z \otimes \gamma_y \\- \lambda_1^c({\bm k}) \tau_0 \otimes \sigma_x \otimes \gamma_y - \lambda_2^c({\bm k}) \tau_z\otimes\sigma_y \otimes \gamma_y  - \lambda_1^s({\bm k}) \tau_0 \otimes \sigma_x \otimes \gamma_x + \lambda_2^s({\bm k}) \tau_z\otimes\sigma_y \otimes \gamma_x + \Delta \tau_y \otimes \sigma_y \otimes \gamma_0. 
    \end{multline}

Here, the $\tau, \sigma, \gamma$ matrices are Pauli matrices acting on the particle-hole (Nambu), spin and orbital degrees of freedom respectively. We have also defined
    \begin{align*}
        t_c({\bm k}) &= \frac{1}{2}\left( t_1 \cos(k_y) +t_2\cos\left(\tfrac{\sqrt{3}}{2}k_x+\tfrac{1}{2}k_y\right)+t_3\cos\left(\tfrac{\sqrt{3}}{2}k_x-\tfrac{1}{2}k_y\right)\right), \\ t_s({\bm k}) &= \frac{1}{2}\left( t_1 \sin(k_y) -t_2\sin\left(\tfrac{\sqrt{3}}{2}k_x+\tfrac{1}{2}k_y\right)+t_3\sin\left(\tfrac{\sqrt{3}}{2}k_x-\tfrac{1}{2}k_y\right)\right), \\ J_i^c({\bm k}) &= -\frac{\alpha_1J_i}{2} \cos(k_y) + \frac{\alpha_1 J_i}{4}\left[\cos\left(\tfrac{\sqrt{3}}{2}k_x+\tfrac{1}{2}k_y\right)+\cos\left(\tfrac{\sqrt{3}}{2}k_x-\tfrac{1}{2}k_y\right) \right] + \frac{\sqrt{3}\alpha_2J_i}{4}\left[\cos\left(\tfrac{\sqrt{3}}{2}k_x+\tfrac{1}{2}k_y\right)-\cos\left(\tfrac{\sqrt{3}}{2}k_x-\tfrac{1}{2}k_y\right) \right], \\ J_i^s({\bm k}) &= -\frac{\alpha_1J_i}{2} \sin(k_y) - \frac{\alpha_1 J_i}{4}\left[\sin\left(\tfrac{\sqrt{3}}{2}k_x+\tfrac{1}{2}k_y\right)-\sin\left(\tfrac{\sqrt{3}}{2}k_x-\tfrac{1}{2}k_y\right) \right] - \frac{\sqrt{3}\alpha_2J_i}{4}\left[\sin\left(\tfrac{\sqrt{3}}{2}k_x+\tfrac{1}{2}k_y\right)+\sin\left(\tfrac{\sqrt{3}}{2}k_x-\tfrac{1}{2}k_y\right) \right], \\ \lambda_1^c ({\bm k}) &= \frac{\lambda}{2} \cos(k_y) - \frac{\lambda}{4}\left[\cos\left(\tfrac{\sqrt{3}}{2}k_x+\tfrac{1}{2}k_y\right)+\cos\left(\tfrac{\sqrt{3}}{2}k_x-\tfrac{1}{2}k_y\right) \right], \\ \lambda_2^c ({\bm k}) &= \frac{\sqrt{3}\lambda}{4}\left[\cos\left(\tfrac{\sqrt{3}}{2}k_x+\tfrac{1}{2}k_y\right)-\cos\left(\tfrac{\sqrt{3}}{2}k_x-\tfrac{1}{2}k_y\right) \right], \\ \lambda_1^s ({\bm k}) &= \frac{\lambda}{2} \sin(k_y) + \frac{\lambda}{4}\left[\sin\left(\tfrac{\sqrt{3}}{2}k_x+\tfrac{1}{2}k_y\right)-\sin\left(\tfrac{\sqrt{3}}{2}k_x-\tfrac{1}{2}k_y\right) \right], \\ \lambda_2^s ({\bm k}) &= \frac{\sqrt{3}\lambda}{4}\left[\sin\left(\tfrac{\sqrt{3}}{2}k_x+\tfrac{1}{2}k_y\right)+\sin\left(\tfrac{\sqrt{3}}{2}k_x-\tfrac{1}{2}k_y\right) \right].
    \end{align*}
\end{widetext}

Using this form of the Hamiltonian, we can easily analyse the symmetries of the model. In particular, we are interested in particle-hole, time-reversal, and chiral symmetry as these define the class to which the topological superconductor belongs \cite{AltlandZirnbauer, TopologyPeriodicTable}.
It can be easily verified that particle-hole symmetry is present via showing that $\tau_x H^*(-{\bm k})\tau_x= -H({\bm k})$, whereas TRS is broken by the altermagnetic exchange terms which violate the condition $H({\bm k})=\sigma_yH^*(-{\bm k})\sigma_y$. There is also no chiral symmetry meaning that the Hamiltonian describes a TSC in class D \cite{AltlandZirnbauer, TopologyPeriodicTable}, and therefore we analyse the Chern number as the topological invariant.

\section*{Phase Diagrams including both orbital symmetries}

\begin{figure}[H]
    \centering
    \includegraphics[width=\linewidth]{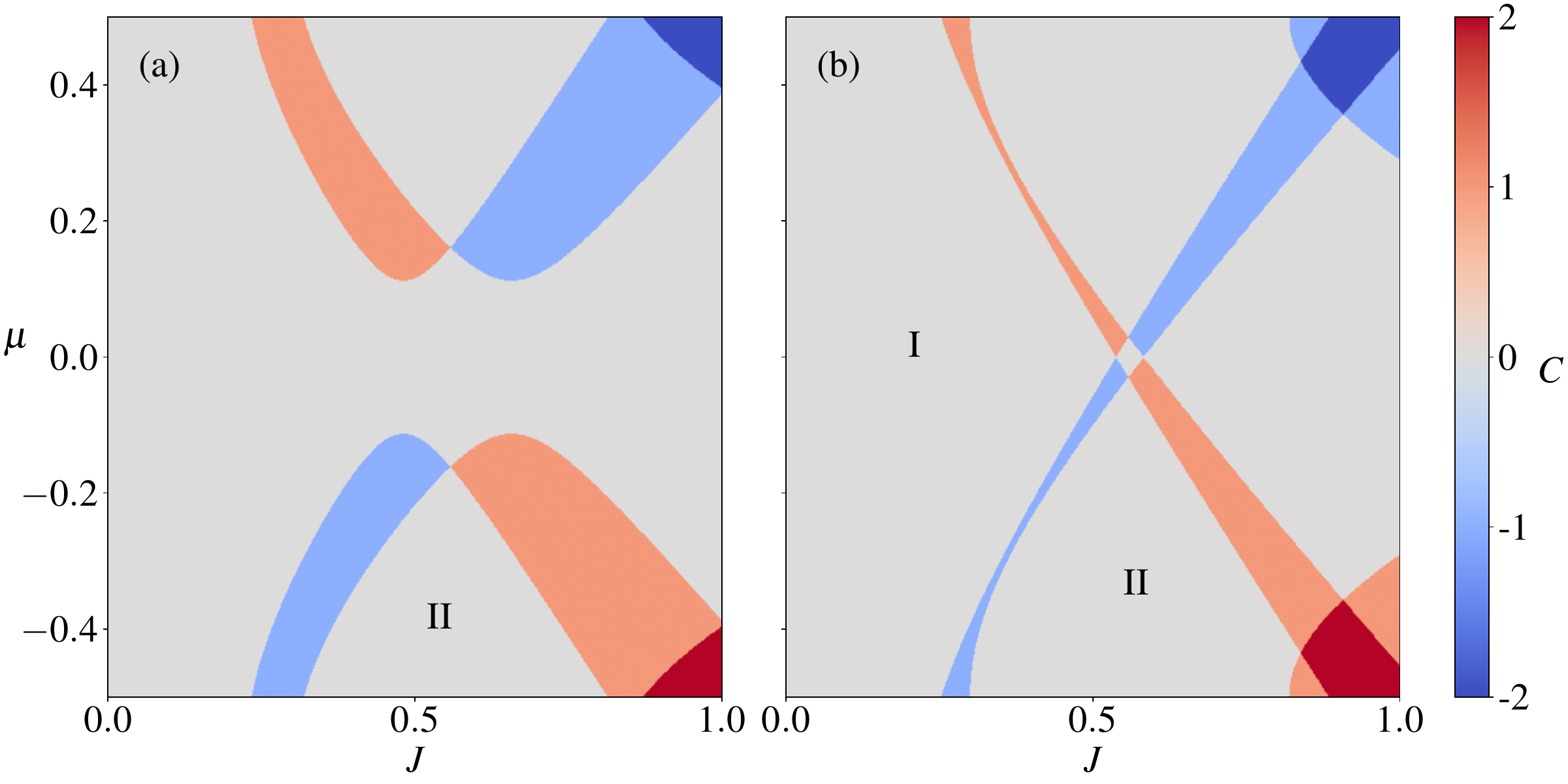}
    \caption{The phase diagrams showing the Chern number for (a): $M=0.1$ and all $t_i=1$, and (b): $M=0$, $t_1=t_3=1$ and $t_2=0.8$. Both of the $d$-wave exchange symmetries are present here ($\alpha_1=\alpha_2=1$ in the Hamiltonian). In both plots the $C=0$ regimes located at the upper and lower central regions are analogous to region II discussed in the main text. The $C=0$ region on the right of (b) contains gapped-out edge modes, as seen in Fig.~5(b) of the main text. In (b) we use a $1001 \times 1001$ grid to discretise the Brillouin zone in the calculation of the Chern number.}
    \label{fig:BothOrbitalsChern}
\end{figure}

Here, we present the phase diagrams when both $d$-orbital symmetries are present. We do this for $M\neq0$ and symmetric hopping (Fig.~\ref{fig:BothOrbitalsChern}(a)) and $M=0$ with symmetric hopping (Fig.~\ref{fig:BothOrbitalsChern}(b)).

In the case of finite $M$, the presence of the $d_{xy}$ symmetry alone does not contribute any new topological physics. However, it's inclusion does affect the phase diagram. In particular is the reappearance of a regime analogous to regime II in Fig~2 of the main text. In this region there are counter-propagating edge modes.

When both $d$-orbital symmetries are present but the asymmetry between the sublattices is caused by asymmetric hopping then the phase diagram looks similar to a combination of those shown in the main text (Fig.~2 and Fig.~5(b)). However the parameter ranges over which certain topological regimes can be observed are narrower. This should be taken into account when modelling and measuring specific materials.

\section*{Robustness of Corner Modes to clusters of vacancies}

\begin{figure}[H]
    \centering
    \includegraphics[width=\linewidth]{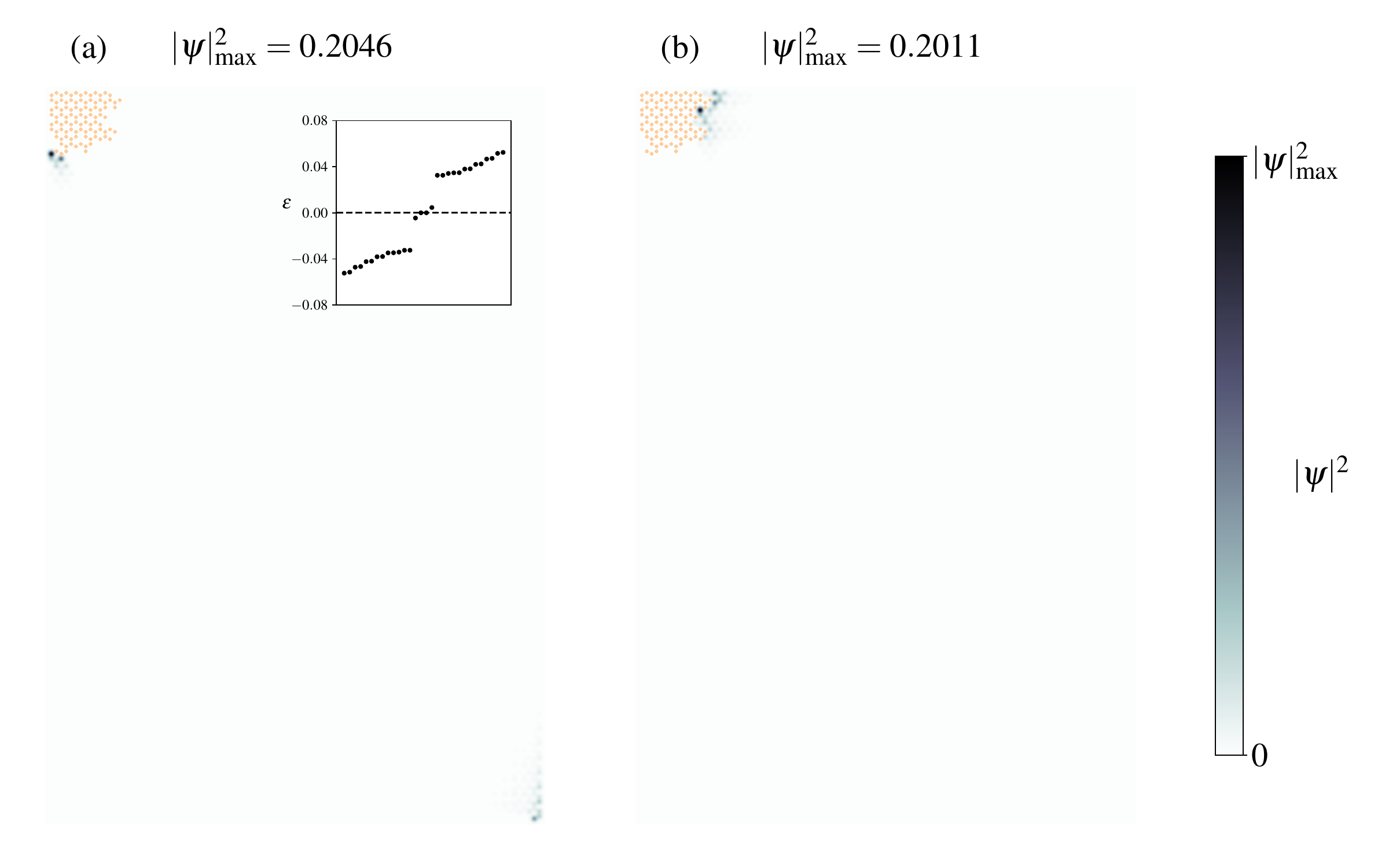}
    \caption{The two lowest in magnitude energy eigenstates when there is cluster of vacancies (shown in orange) in the top left corner. The system size is $101 \times 100$ in the absence of vacancies. (a) shows that the zero-energy corner modes are robust. The inset and (b) additionally show that there are additional modes close to zero energy, localised near the disordered corner.}
    \label{fig:Cluster_Corner1}
\end{figure}

In this section, we present results that the zero-energy corner modes are robust to disordered corners. To model this we use a cluster of vacancies located at the corners where the modes are present. All parameters are the same as in Fig.~2 of the main text with $J=0.8, \mu=-0.1$. The figure, Fig.~\ref{fig:Cluster_Corner1}, shows the vacancies occurring in the top left corner. Similar results are found if the disordered corner is the bottom right one.

\section*{Additional Disorder Configurations - Corner Modes}

\begin{figure}[H]
    \centering
    \includegraphics[width=\linewidth]{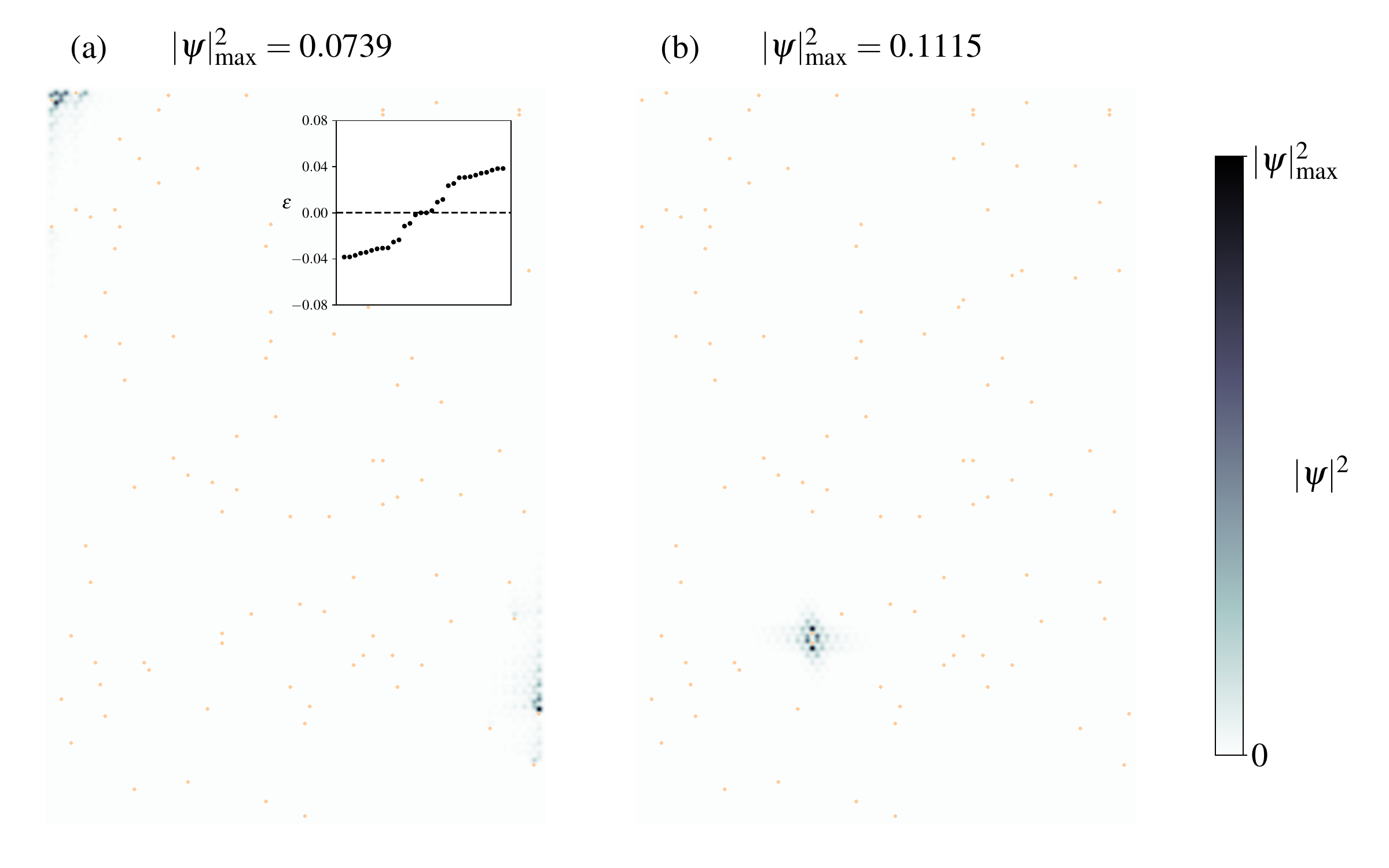}
    \caption{The probability distribution for the two lowest absolute energy eigenstates, when $1\%$ of the sites in the $101\times 100$ lattice are vacancies. In (a), it is shown that the spatial extent of the corner modes can be partially affected by the presence of nearby vacancies. There are also additional modes with energy close to zero - see the inset and (b).}
    \label{fig:Additional_Configs2}
\end{figure}

\begin{figure}[H]
    \centering
    \includegraphics[width=\linewidth]{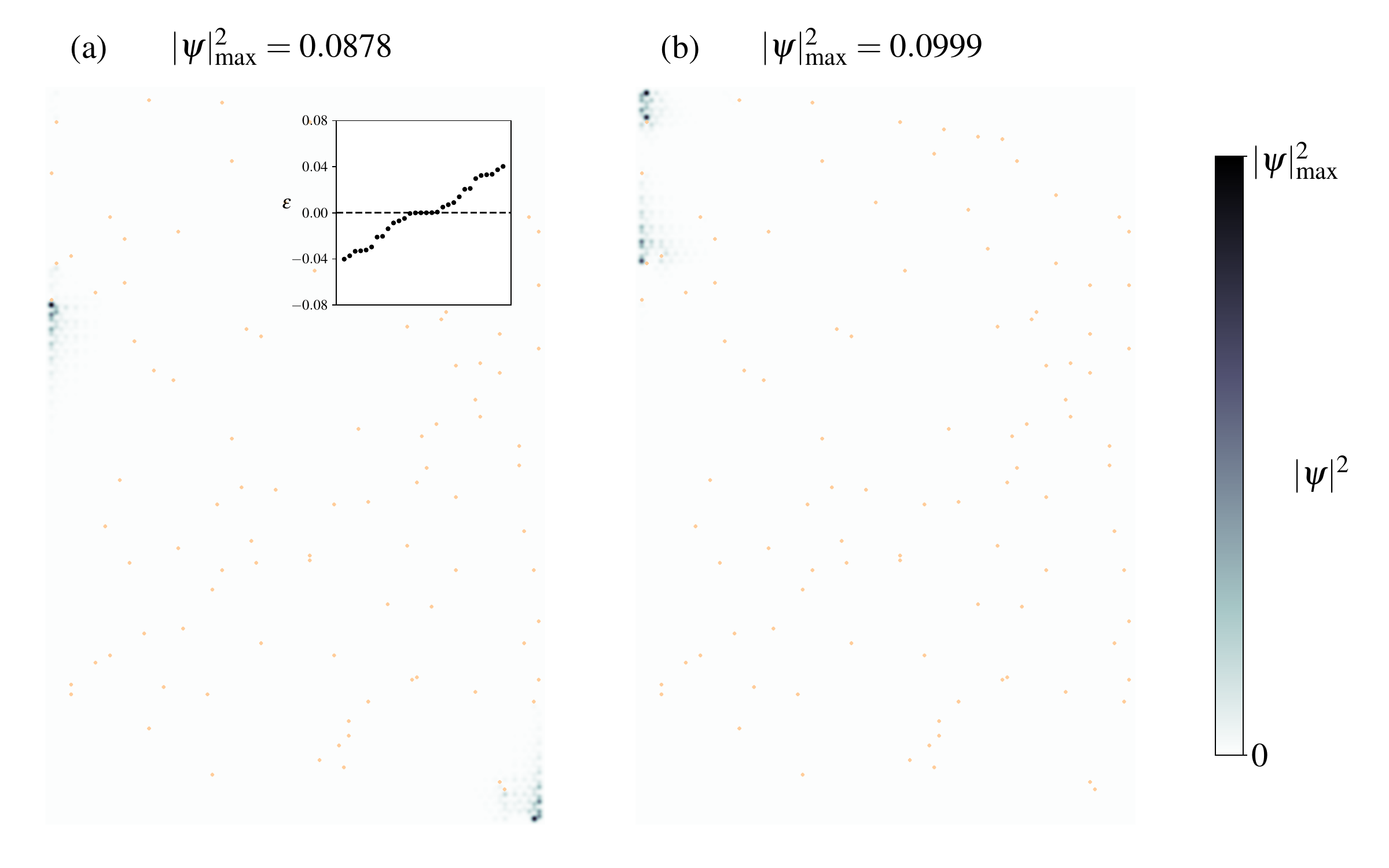}
    \caption{The two lowest absolute energy states for $1\%$ disorder in a $101 \times 100$ lattice. The modes pinned to the disorder can be near degenerate with the corner modes (see inset). (a) and (b) show these (near) degenerate modes. Therefore the modes may no longer look like they are localised to the corners only.}
    \label{fig:Additional_Configs3}
\end{figure}

Here, we provide further evidence that whilst the corner modes may survive the presence of disorder, the potential detection of them is dependent on the microscopic details. This is due to the existence of other low energy modes that are localised near vacancies. The parameters used are again the same as Fig.~2 of the main text with $J=0.8, \mu=-0.1$. In Fig.~\ref{fig:Additional_Configs2}, we show that the spatial extent of the corner modes can be modified by the presence of vacancies (we use a $1\%$ disorder level), potentially hampering their detection. A further complication is shown in Fig.~\ref{fig:Additional_Configs3}, where we show that the impurity modes can be nearly degenerate with the zero-energy corner modes. This can give the perception that the modes are no longer localised to the corners.

